\documentclass[12pt]{article}
\usepackage{amsfonts,amsmath,amssymb,mathrsfs,url}

\title{What Does the Free Will Theorem Actually Prove?}
\author{
Sheldon Goldstein\footnote{Departments of Mathematics, Physics and
     Philosophy, Rutgers University,
     110 Frelinghuysen Road, Piscataway, NJ 08854-8019, USA.
     E-mail: oldstein@math.rutgers.edu},
Daniel V. Tausk\footnote{Departamento de Matem\'atica,
    Universidade de S\~ao Paulo,
    Rua do Mat\~ao 1010, Cidade Universit\'aria,
    CEP 05508-090 S\~ao Paulo - SP, Brazil.
    E-mail: tausk@ime.usp.br},\\
Roderich Tumulka\footnote{Department of Mathematics,
     Rutgers University,
     110 Frelinghuysen Road, Piscataway, NJ 08854-8019, USA.
     E-mail: tumulka@math.rutgers.edu},
\ and Nino Zangh\`\i\footnote{Dipartimento di Fisica dell'Universit\`a
     di Genova and INFN sezione di Genova, Via Dodecaneso 33, 16146
     Genova, Italy. E-mail: zanghi@ge.infn.it}
}
\date{May 28, 2009}

\newcommand{\PPP}{\mathbb{P}}
\newcommand{\RRR}{\mathbb{R}}

\newcommand{\be}{\begin{equation}}
\newcommand{\ee}{\end{equation}}

\newcommand{\achoice}{\mathfrak a}
\newcommand{\bchoice}{\mathfrak b}
\newcommand{\Aoutcome}{O_A}
\newcommand{\Boutcome}{O_B}

\begin{document}
\maketitle

Conway and Kochen have presented a ``free will theorem'' \cite{CK05,CK09} which they claim shows that ``if indeed we humans have free will, then [so do] elementary
particles.'' In a more precise fashion, they claim it shows that for certain quantum experiments
in which the experimenters can choose between several options, no deterministic or stochastic model can account for the
observed outcomes without violating a condition ``MIN'' motivated by relativistic symmetry. We point out that for
stochastic models this conclusion is not correct, while for deterministic models it is not new.

In the way the free will theorem is formulated and proved, it only concerns deterministic models. But Conway and Kochen have argued \cite{CK05,CK07,CK09,CK09b} that ``randomness can't help,'' meaning that stochastic models are
excluded as well if we insist on the conditions ``SPIN'', ``TWIN'', and ``MIN''. We point out a mistake in their argument. Namely, the theorem is of the form
\be\label{claim1}
\text{deterministic model with SPIN \&\ TWIN \&\ MIN $\Rightarrow$ contradiction}\,,
\ee
and in order to derive the further claim, which is of the form
\be\label{claim2}
\text{stochastic model with SPIN \&\ TWIN \&\ MIN $\Rightarrow$ contradiction}\,,
\ee
Conway and Kochen propose a method for converting any stochastic model into a deterministic one \cite{CK05}:
\begin{quotation}
``let the stochastic element \dots be a sequence
of random numbers (not all of which need be
used by both particles). Although these might
only be generated as needed, \textbf{it will plainly make
no difference to let them be given in advance}.'' [emphasis added]
\end{quotation}
In this way, \eqref{claim2} would be a corollary of \eqref{claim1} if the conversion preserved the properties
SPIN, TWIN, and MIN. However, Conway and Kochen have neglected to check whether they are preserved, and indeed,
as we will show, the conversion preserves only SPIN and TWIN but not MIN. We do so by exhibiting a simple
example of a stochastic model satisfying SPIN, TWIN, and MIN. As a consequence, no method of conversion of
stochastic models into deterministic ones can preserve SPIN, TWIN, and MIN. More directly, our example shows
that \eqref{claim2} is false. Contrary to the emphasized part of the above quotation, letting the randomness be given
in advance makes a big difference for the purpose at hand.

The relevant details are as follows. The reasoning concerns a certain experiment in which, after a preparation procedure,
two experimenters ($A$ and $B$), located in space-time regions that are spacelike separated, can each choose
between several options for running the experiment.
We denote by $\achoice$ (resp., by $\bchoice$) the choice of $A$ (resp., of $B$) and by $\Aoutcome$ (resp., $\Boutcome$) the outcome of $A$
(resp., of $B$). The data collected from this experiment can be represented by a joint probability
distribution $\PPP_{\achoice\bchoice}(\Aoutcome,\Boutcome)$ for the outcomes $(\Aoutcome,\Boutcome)$ that depends on the choices
$(\achoice,\bchoice)$. Experimenter $A$ chooses
$\achoice=(x,y,z)$ from a certain set of 40 orthonormal bases of $\RRR^3$, $B$ chooses $\bchoice=w$ from a certain set of 33 unit
vectors in $\RRR^3$. SPIN asserts that the outcome $\Aoutcome$ obtained by $A$ is always one of the triples 110, 101, or 011, and the outcome $\Boutcome$ obtained by $B$ is either 0 or 1. TWIN asserts that whenever $w=x$
(resp., $w=y$/$w=z$) then $\Boutcome$ coincides with the first (resp., second/third) digit of $\Aoutcome$.
Quantum mechanics predicts data $\PPP_{\achoice\bchoice}(\Aoutcome,\Boutcome)$ that satisfy SPIN and TWIN, given
explicitly in Table~\ref{QMpredict}.

\begin{table}[h]
\begin{center}
\begin{tabular}{r|cc}
$\PPP_{\achoice\bchoice}$ & $\Boutcome=0$ & $\Boutcome=1$\\
\hline
\vrule height 13pt depth 0pt width 0pt$\Aoutcome=011$ & $\tfrac13(w\cdot x)^2$ & $\tfrac13[1-(w\cdot x)^2]$\\[3pt]
$\Aoutcome=101$ & $\tfrac13(w\cdot y)^2$ & $\tfrac13[1-(w\cdot y)^2]$\\[3pt]
$\Aoutcome=110$ & $\tfrac13(w\cdot z)^2$ & $\tfrac13[1-(w\cdot z)^2]$
\end{tabular}
\end{center}
\caption{Joint probability distribution of outcomes as predicted by quantum mechanics, with $\cdot$ denoting the scalar product of vectors in $\RRR^3$}
\label{QMpredict}
\end{table}

A \emph{stochastic model\/} for the data $\PPP_{\achoice\bchoice}(\Aoutcome,\Boutcome)$
means, for the purpose at hand, a probability measure $\PPP^\Lambda$
(that does not depend on $\achoice$ and $\bchoice)$ on some measurable space $\Lambda$ and, for each
$\lambda\in\Lambda$ and $\achoice$ and $\bchoice$, a probability measure
$\PPP_{\achoice\bchoice}(\Aoutcome,\Boutcome|\lambda)$ on the set $\{110,101,011\}\times\{0,1\}$ of possible outcomes
such that, when $\lambda$ is averaged over with $\PPP^\Lambda$, the data $\PPP_{\achoice\bchoice}(\Aoutcome,\Boutcome)$
are obtained:
\be\label{agree}
\PPP_{\achoice\bchoice}(\Aoutcome,\Boutcome)=\int_\Lambda\PPP_{\achoice\bchoice}(\Aoutcome,\Boutcome|\lambda)\,\mathrm d\PPP^\Lambda(\lambda).
\ee
A \emph{deterministic model\/} for the data $\PPP_{\achoice\bchoice}(\Aoutcome,\Boutcome)$ is a stochastic model such that each $\PPP_{\achoice\bchoice}(\Aoutcome,\Boutcome|\lambda)$ is supported by a single outcome, i.e., one for which there are functions $\theta_A$ and $\theta_B$ such that:
\be
\PPP_{\achoice\bchoice}\big(\Aoutcome=\theta_A(\achoice,\bchoice,\lambda),\ \Boutcome=\theta_B(\achoice,\bchoice,\lambda)|\lambda\big)=1,
\ee
for all $\achoice$, $\bchoice$, and $\lambda$.

The MIN condition is formulated in a somewhat vague way \cite{CK09}:
\begin{quotation}
``The MIN Axiom: Assume that the experiments performed by $A$ and $B$ are space-like separated.
Then experimenter $B$ can freely choose any one of
the 33 particular directions $w$, and [$\Aoutcome$]
is independent of this choice. Similarly and independently, $A$ can freely choose any one of the 40 triples $x, y , z$, and [$\Boutcome$] is independent of that choice.''\footnote{Here and in the following quotation, we have adapted the notation by putting [$\Aoutcome$] for ``$a$'s response''
and [$\Boutcome]$ for ``$b$'s response.''}
\end{quotation}
What does MIN mean for a deterministic model? According to
Conway and Kochen \cite{CK09}:
\begin{quotation}
``It is possible to give a more precise form of MIN
by replacing the phrase `[$\Boutcome$] is
independent of $A$'s choice' by `if [$\Aoutcome$] is
determined by $B$'s choice, then its value does not
vary with that choice.'{}''
\end{quotation}
That is, MIN asserts that the function $\theta_A$ does not depend on $\bchoice$ and the function $\theta_B$ does
not depend on $\achoice$:
\be\label{MINdet}
\theta_A(\achoice,\bchoice,\lambda) = \theta_A(\achoice,\lambda)\,,\quad
\theta_B(\achoice,\bchoice,\lambda) = \theta_B(\bchoice,\lambda)\,.
\ee
What does MIN mean for a stochastic model? Conway and Kochen do not say precisely, as the above quotation deals only
with the case of a deterministic model (``if [$\Aoutcome$] is determined by $B$'s choice''), but the most reasonable
interpretation is a condition known as \emph{parameter independence} \cite{Jar84,Shi84}:
\textit{for any given $\lambda$, the distribution of $\Aoutcome$ does not depend on $\bchoice$, and the distribution
of $\Boutcome$ does not depend on $\achoice$}:
\be\label{PI}
\PPP_{\achoice\bchoice}(\Aoutcome|\lambda) = \PPP_{\achoice}(\Aoutcome|\lambda)\,, \quad
\PPP_{\achoice\bchoice}(\Boutcome|\lambda) = \PPP_{\bchoice}(\Boutcome|\lambda)\,.
\ee
Note that for deterministic models \eqref{PI} is the same as \eqref{MINdet}.

An example of a stochastic model satisfying SPIN, TWIN, and MIN (understood as \eqref{PI}) is obtained from ``rGRWf'',
the relativistic Ghirardi--Rimini--Weber theory with flash ontology \cite{Tum04,Tum07}, but much simpler examples are possible. As a second example, one may simply take $(\Lambda,\PPP^\Lambda)$ to be the trivial probability space containing just one element (so that $\lambda$ is a constant and can be ignored). Then, according to the definition of stochastic models, the data themselves form a stochastic model.
That is, take $\PPP_{\achoice\bchoice}(\Aoutcome,\Boutcome|\lambda)=\PPP_{\achoice\bchoice}(\Aoutcome,\Boutcome)$ as given by Table~\ref{QMpredict}. We know that this stochastic model satisfies SPIN and TWIN, and it also satisfies \eqref{PI}, since, for all $\achoice$ and $\bchoice$, the marginal distribution of $\Aoutcome$
is uniform and the marginal distribution of $\Boutcome$ gives probability 1/3 to 0 and 2/3 to 1. As a third (and even simpler) example, let us drop the requirement \eqref{agree} that the stochastic model agrees with the data predicted by quantum mechanics and focus just on satisfying SPIN, TWIN and MIN. Take $(\Lambda,\PPP^\lambda)$ to
be trivial as before. If $\bchoice=w$ coincides with coordinate $x$ (resp., $y$/$z$) of $\achoice$ then let
$\PPP_{\achoice\bchoice}(\Aoutcome,\Boutcome|\lambda)$ give probability $1/3$ to each of $(110,1)$, $(101,1)$, $(011,0)$ (resp., to each of $(110,1)$,
$(101,0)$, $(011,1)$/to each of $(110,0)$, $(101,1)$, $(011,1)$) and probability zero to the other three possible
values of $(\Aoutcome,\Boutcome)$. If $w$ coincides with none of $x,y,z$ then let $\PPP_{\achoice\bchoice}(\Aoutcome,\Boutcome|\lambda)$ give
probability 1/9 to each of $(110,0)$, $(101,0)$, $(011,0)$ and probability 2/9 to each of $(110,1)$, $(101,1)$,
$(011,1)$. Then SPIN and TWIN are obviously true, and \eqref{PI} is true because the marginal distributions of $\Aoutcome$ and $\Boutcome$ are the same as in the previous example.

To illustrate explicitly why \eqref{PI} breaks down when putting all randomness in the past, let us consider a specific conversion method of stochastic models into deterministic ones that Conway and Kochen have proposed \cite{CK09} in response to earlier criticisms of their claims concerning the viability of rGRWf \cite{Tum07}:
\begin{quotation}
``we can easily deal with the dependence of the
distribution of flashes on the external fields $F_A$ [= $\achoice$]
and $F_B$ [= $\bchoice$], which arise from the two experimenters'
choices of directions $x, y , z$, and $w$. There are
$40 \times 33 = 1320$ possible fields in question. For
each such choice, we have a distribution $X (F_A, F_B )$
of flashes, i.e., we have different distributions
$X_1 , X_2 , \dots, X_{1320}$. Let us be given `in advance'
all such random sequences, with their different
weightings as determined by the different fields. Note that for this to be given, nature does not have to know
in advance the actual free choices $F_A$ (i.e., $x, y, z$) and $F_B$ (i.e., $w$) of the experimenters. Once the
choices are made, \textbf{nature need only refer to the relevant random sequence $\mathbf{X_k}$ in order to emit
the flashes in accord with rGRWf}.'' [emphasis added]
\end{quotation}
The problem here is that the deterministic model obtained from this method of conversion manifestly violates
MIN because if nature were to follow the recipe suggested in the emphasized part of the quotation above then she
would have to use the value of $k=k(x,y,z,w)$ depending on  both experimenters' choices, $\achoice=(x,y,z)$, and $\bchoice=w$, in order to produce
any of the outcomes $\Aoutcome$, $\Boutcome$.

The conclusion that there are some predictions of quantum theory
that cannot be obtained by a deterministic model satisfying parameter independence is not new. As noted by
Jarrett in 1984 \cite{Jar84}, for a stochastic model \emph{Bell's locality condition}
\cite{Bell64,Bell87b}
\be\label{locality}
\PPP_{\achoice\bchoice}(\Aoutcome,\Boutcome|\lambda) = \PPP_{\achoice}(\Aoutcome|\lambda) \, \PPP_{\bchoice}(\Boutcome|\lambda)
\ee
is (straightforwardly) equivalent to the conjunction of parameter independence \eqref{PI}
and another condition known as \emph{outcome independence},
\be\label{OI}
\PPP_{\achoice\bchoice}(\Aoutcome|\Boutcome,\lambda) = \PPP_{\achoice\bchoice}(\Aoutcome|\lambda)\,,\quad
\PPP_{\achoice\bchoice}(\Boutcome|\Aoutcome,\lambda) = \PPP_{\achoice\bchoice}(\Boutcome|\lambda)\,.
\ee
For a deterministic model, \eqref{OI} is always trivially satisfied, as the distributions
$\PPP_{\achoice\bchoice}(\Aoutcome|\lambda)$ and $\PPP_{\achoice\bchoice}(\Boutcome|\lambda)$ each assign probability 1 to a single outcome
so any further information (like the other outcome) is redundant.
Thus, for a deterministic model, parameter independence is equivalent to locality, which Bell showed in
1964 \cite{Bell64} to be incompatible with some predictions of quantum mechanics.
Therefore, deterministic models in agreement with quantum predictions must violate parameter independence.
Even from the very same experiment as considered by Conway and Kochen, this conclusion was derived before
in \cite{S,HR,BS,Elby} and \cite[Section 4.2.1]{Hemmick}, in \cite{Hemmick} using only SPIN and TWIN.

It has been suggested to one of us (R.T.) by Simon Kochen that our understanding of MIN  is too weak, that MIN should be regarded as requiring that the actual outcome  itself of $A$ be independent of $B$'s choice, and not just its probability distribution. We are unable to see why this is a reasonable requirement for a stochastic theory---or even what exactly it should mean. Be that as it may, the existence of the examples described here demonstrates that any such variant of MIN for a stochastic model would either be unreasonable (or worse) or would fail to be preserved under conversion of the model to a deterministic one.

\bigskip

\noindent\textit{Acknowledgements.}
S.~Goldstein was supported in part by NSF Grant
DMS-0504504.
N.~Zangh\`\i\ was supported in part by Istituto Nazionale di Fisica
Nucleare.

\end{document}